\documentclass[aps,pre,preprint,eqnum]{revtex4}
\usepackage{graphicx}
\usepackage{graphics}
\usepackage{psfig}

\begin{document}
\title{Rabi type oscillations in damped single 2D-quantum dot}

\author{Madhuri Mukhopadhyay${^a}$, Ram Kuntal Hazra${^a}$,
Manas Ghosh${^b}$, S. Mukherjee${^a}$, S.P. Bhattacharyya${^a}$ {\footnote {e-mail address:
pcspb@iacs.res.in}}\footnote{Phone : (91)(33)2473 4971, 3372, 3073, 5374} \footnote{Fax : (91)(33)2473 2805}}

\affiliation{$^{a}$Department of Physical Chemistry, Indian
Association
for the Cultivation of Science Jadavpur, Kolkata, India \\
$^{b}$ Department of Chemistry, Physical Chemistry section, Visva
Bharati University Santiniketan, Birbhum, India}

\begin{abstract} We present a quantized model of harmonically
confined dot atom with inherent damping in the presence of a
transverse magnetic field. The model leads to a non hermitian
Hamiltonian in real coordinate. We have analytically studied the
effects that damping has on the $Rabi$ type oscillations of the
system. The model explains the decoherence of Rabi oscillation in a
Josephson Junction.

\vspace{0.8cm}


\textbf{keywords}: damped quantum dot, quantization of damping, Rabi oscillation.

\end{abstract}

\vspace{0.3cm}

\maketitle

\section{Introduction}

Rabi oscillation \cite{Rabi37} is one of the fundamental observations in light
matter interaction that occurs coherently and nonlinearly
\cite{Zubairy} and which has no classical analogue. The generation
of coherent superposition of quantum states using ultra short laser
pulses and the subsequent decoherence due to some inherent damping
or interaction with the environment is of great interest especially
in semiconductor quantum dots due to the prospect of future
applications \cite{YuA03, Wallraff04, Vion02} in quantum information
processing and making novel laser devices \cite{Brien07}. Rabi
oscillations using excitons in single quantum dots
\cite{Ghosh05,Fussel07,Hazra07,Rao07} have been studied successfully
by different groups in the past few years
\cite{Binder99,Steie01,Martini02,Wallraff05}. Control of the
decoherence of Rabi oscillation in quantum dot, the mechanism of
which is still a matter of investigation, has attracted wide
attention \cite{Zrenner02,Kosugi05}. There has been rapid progress
in experimental control of dephasing of coherent states in quantum
dots \cite{Villas05}. On the contrary, theoretical studies on
quantum dots leading to the dynamics in the presence of damping are
scarce. Till date theoretical studies in dots have been done mainly
on the basis of damping that has been introduced phenomenologically.
To the best of our knowledge, no quantum theoretical model of the
dot has been developed with the inherent damping incorporated in the
model.

 We develop a model of a damped quantum
dot going beyond the phenomenological description used so far. The
model, we believe provides some insight into the Rabi dynamics. Our
analytical results lead to an understanding of the experimental
observation of decoherence in Josepshon Junction \cite{Yu02}.

\section{Model}
We show, in what follows, that the damped one electron dot can be
described by the eigenstates of a quantum Hamiltonian H that is non
hermitian. The artificial atom that we have modeled is composed of a
single electron confined in 2-D by harmonic potential with some
inherent damping and a homogeneous magnetic field applied normal to
the confinement plane.
 Let us start with the classical equation
of motion of the damped harmonic oscillator which reads
\begin{equation}
\label{eq1} m_e\ddot{\overrightarrow r}+\gamma\dot{\overrightarrow
r}+k{\overrightarrow r}=0
\end{equation}
where k is the harmonic force constant and $\gamma$ is the damping
constant and $m_e$ is oscillator mass. The system described by
equation (1) is known to have a time dependent Lagrangian and
Hamiltonian \cite{gold,ray79,Riewe}. There have been many attempts
to quantize the damped linear oscillator \cite{Dekker77, Dekker81,
Harris90} but a completely satisfactory solution been elusive. The
stumbling block has been the lack of a time independent Hamiltonian
formalism. Recently, however such a formalism has been proposed
making a definite progress \cite{latimar05,chandrasekhar07}. We
proposed a different strategy that brings a non hermitian
Hamiltonian formalism. From equation (1) we start by noting that it
is immediately possible to write down the Euler-Lagrange equation
for the dissipative system by defining a velocity dependent force
$F_{gen}$ and setting
\begin{eqnarray}
\frac{d}{dt} \left(\frac{\partial L}{\partial \dot{r}}
\right)-\frac{\partial L}{\partial r} = F_{gen}
\end{eqnarray}
where $F_{gen}$ is defined as the negative derivative of Rayleigh
dissipative function $'f'$ with respect to $\dot{r}$ \cite{gold}.
\begin{equation}
\label{eq5} F_{gen}=-\frac{\partial}{\partial \dot{r}}(f)
\end{equation}
 $f$ is determined by the damping constant $\gamma$ and the
velocity $(\dot{r})$ as follows:
\begin{equation}
f= \frac{1}{2}\gamma\dot{r}^{2}
\end{equation}
Equations (3) and (4) suggest that the time dependent damping force
$(F_d)$ is linearly related to the velocity:
\begin{equation}
F_d=-\gamma \dot{\overrightarrow{r}}
\end{equation}
With equation (3) the Euler Lagrange equation (2) now reads
\begin{equation}
\frac{d}{dt} \left(\frac{\partial L}{\partial \dot{r}}
\right)-\frac{\partial L}{\partial r}+ \frac{\partial f}{\partial
\dot{r}}=0
\end{equation}
Equation (6) requires that the Lagrangian L is chosen as
\begin{equation}
L=\frac{1}{2}m_e \dot{r}^2+\gamma r\dot{r}-\frac{1}{2}kr^{2}
\end{equation}
Clearly the Lagrangian of equation (7) is consistent with the
equation of motion of the damped harmonic oscillator equation (1).
Since the momentum $p=\frac{\partial L}{\partial \dot{r}}$ the
modified momentum for the damped harmonic oscillator becomes
\begin{equation}
p=m_e\dot{\overrightarrow{r}}+\gamma\overrightarrow{r}
\end{equation}
Let the damped oscillator have a charge 'q' and let it experience an
electric field (E) and a transverse magnetic field (B). The Lorentz
force acting on it is
\begin{equation}
F=q \left[E + \frac{1}{c}(v\times B)\right]=q\left[-\nabla\phi
-\frac{1}{c}\left( \frac{\partial A}{\partial t}\right) +
\frac{1}{c}(v\times B)\right]
\end{equation}
where $E=-\nabla \phi$ ($\phi = $scalar potential) and $B= \nabla
\times A $ (A= Vector potential).\\ The electric and magnetic fields
bring in additional terms in the Lagrangian ($L=\bar{L}$, say) where
\begin{equation}
\bar{L}=\frac{1}{2}m_e \dot{r}^2+\gamma r\dot{r}-q\phi +
\frac{q}{c}\overrightarrow{A}\dot{\overrightarrow r}
\end{equation}\\,
where $q\phi=\frac{1}{2}kr^{2}$, scalar potential. The modified
momentum $(\bar{p})$ for the system (described by $\bar{L}$)
\begin{equation}
\bar{p} = m_e\dot{\overrightarrow r}+\gamma {\overrightarrow r}+
\frac{q}{c}\overrightarrow{A}.
\end{equation}
The modified momentum  $\bar{p}$ leads to the Hamiltonian
$(\bar{H})$ of the system represented by a single carrier electron
in a damped quantum dot as follows:
\begin{equation}
\bar{H}=\frac{1}{2m_e}\left[(m_e \dot{\overrightarrow r}+\gamma
{\overrightarrow r}+ \frac{q}{c}\overrightarrow{A})\cdot (m_e
\dot{\overrightarrow r}+\gamma {\overrightarrow r}+
\frac{q}{c}\overrightarrow{A})\right]+ q\phi
\end{equation}
Taking the cyclotron frequency $\omega_{c}=\frac{qB}{m_{e}c}$,the
confinement potential $q\phi =\frac{1}{2}m_e
\omega_{0}^{2}(x^2+y^2)$ and replacing the classical operators by
their respective quantum analogues, the quantum mechanical
Hamiltonian of the system in Cartesian coordinates becomes
\begin{eqnarray}
\bar{H} = &-& \frac{\hbar^2}{2m_e}\left(\frac{\partial^2}{\partial
x^2}+\frac{\partial^2}{\partial y^2} \right) -
\frac{i\hbar\gamma}{m_e}\left(1+x\frac{\partial}{\partial x}+y
\frac{\partial}{\partial y}\right)- \frac{i\hbar\omega_c}{2}
\left(-y\frac{\partial}{\partial x} +x\frac{\partial}{\partial
y}\right)\nonumber\\
&+& \frac{\gamma^2}{2m_e}(x^2 + y^2) +
\frac{m_e}{8}\omega_c^2(x^2+y^2)+\frac{1}{2}m_e\omega_0^2(x^2+y^2).
\end{eqnarray}
 Transforming from Cartesian to polar coordinates the Hamiltonian changes to
\begin{eqnarray}
H=-\frac{\hbar^2}{2m_e}\left(\frac {\partial^2}{\partial
r^2}+\frac{1}{r}\frac{\partial}{\partial
r}+\frac{1}{r^2}\frac{\partial^2}{\partial\phi^2}\right)-
\frac{i\hbar\gamma}{m_e}\left(1+r\frac{\partial}{\partial
r}\right)-\frac
{i\hbar\omega_c}{2}\left(\frac{\partial}{\partial\phi}\right)+\Omega_d^2r^2
\end{eqnarray}
where $\Omega_d^2= \frac{1}{2}m_e\left[\frac
{\omega_c^2}{4}+\frac{\gamma^2}{m_e^2}+\omega_0^2\right]$.\\
H is manifestly non-hermitian.
 H may be thought of as defining a set of eigenstates
$\psi_{n,l}(r,\phi)$ with complex energy $E_{n,l}$ if we assume that
H obeys the energy eigenvalue equation
\begin{equation}
H\psi_{n,l}(r,\phi)=E_{n,l}\psi_{n,l}(r,\phi)
\end{equation}

A straight forward series solution of equation (15) (Appendix-A)
leads to the quantized energy eigenvalues of the damped dot:
\begin{equation}
E_{n,l}=  \frac{\omega_c l}{2}+(2n+l+1)\Omega -i
\gamma(2n+l+1)\Omega
\end{equation}where `\emph{n}' and `\emph{l}' are principal,
 and angular momentum
quantum numbers, respectively and $ \Omega^2= \left[\frac
{\omega_c^2}{4}+\frac{\gamma^2}{m_e^2}+\omega_0^2\right]$.\\The
energy is clearly complex and the imaginary part of it is related to
the dissipating energy which is given by
\begin{eqnarray}
\Gamma_{n,l} =-\gamma(2n+l+1)\Omega
\end{eqnarray}

Thus, starting from the classical equation of motion of the damped
harmonic oscillator quantization has been carried out through a
Lagrange-Hamiltonian formalism, where the Hamiltonian is
non-Hermitian \cite{Dekker75,Wartak89} as expected for a non
conservative system \cite{Razavy76,Somnath89}. We have described the
system in terms of real positional coordinates in contrast with
attempts to handle the problem in terms of complex coordinate
\cite{Dekker75}.

Thus, proceeding with the assumption that the system described by
the non-Hermitian Hamiltonian of equation (13)
 satisfies time-independent Schr$\ddot{o}$dinger equation
$H\psi=E\psi$ \cite{Dekker75}, we have obtained all the quasi energy eigenstates. \\
\begin{eqnarray}
\psi_{n,l}(r, \phi)=\frac {C}{2\sqrt{\pi}}e^{-\frac {\Omega^2
r^2}{2}}r^l L_n^{|l|}
\end{eqnarray}
where $L_n^{|l|}$ is the Laguerre series and C is the normalization
constant.
 In the absence of damping these states merge into Fock-Darwin
energy spectrum \cite{Fock28,Darwin30}, while the presence of
damping makes the energy levels quasi stationary. The important
outcome is that for a known $\omega_0$ and $\omega_c$ comparison of
the energy separation between two states as observed from experiment
and obtained from the expressions with and without damping can lead
to the realization of the intrinsic damping coefficient of a dot
system. For damped dot system the energy states are shifted from the
energy levels without damping and the shifts are more pronounced for
stronger damping whereas for greater effective mass  of the carrier
electron the effect of damping is somewhat quenched. Since the
non-hermitian Hamiltonian obtained for the damped dot has complex
eigenvalues that correlate with the energy eigenvalues of the dot in
the limit of zero damping, it could be interesting to investigate
the dynamics of the damped dot in response to perturbation by laser
light.

\section{Dynamics of Damped Quantum Dot:}

Let us consider the time-dependent Schr$\ddot{o}$dinger equation for
the complex energy eigen states of H;
\begin{equation}
\nonumber i\hbar\frac{\partial\Psi_{n,l}(r,t)}{\partial
t}=(E_R-i\Gamma)_{n,l}\Psi_{n,l}(r,t),
\end{equation}
The corresponding wave function is decaying and the probability
P(r,t) is proportional to $ {|\psi_{n,l}(r,0)|}^2e^{\frac{-2\Gamma
t}{\hbar}}$. The exponential function accounts for the exponential
fall-off of the amplitude with time, the first factor being the the
amplitude of the initial state which is now damped. The intrinsic
life time $\tau_{n,l} = \frac{1}{\Gamma}_{n,l}$ of these
quasi-stationary states are therefore determined by the damping
coefficient and the quantum number characterizing the states.

We now consider the two energy levels ($g$  and  $e$) of the damped
quantum dot system,  the two states are designated as $\psi_g$  and
$\psi_e$ are assumed to be well separated from all other states. The
system interacts with a laser of frequency $\omega_L$ and
$\omega_a\equiv\omega_{eg}$ is the resonance frequency (Fig.1). The
effect of perturbation produced by the laser can be treated
semiclassically using the eigenfunctions of the damped dot
$\Psi_{n,l}$ as zeroth order wave function. The perturbed
Hamiltonian is partitioned into $H_0$ and V, where the unperturbed
dot Hamiltonian $H_0$ is given by the equation (13) and the
perturbation in the dipole approximation: $\overrightarrow{V}=
e.rE_0 \cos(\omega_Lt)$.

We may now consider the semiclassical perturbation treatment based
on the damped wave functions of the damped dot already obtained. The
time-dependent Schrodinger equation for the perturbed system is
\begin{eqnarray}
|\dot{\Psi}(r,t)\rangle=-\frac{i}{\hbar}H(r, t)|\Psi(r,t)\rangle
\end{eqnarray}
while the solution is (k=e,g)
\begin{equation}
\Psi(r,t)= \sum_k C_k(t)\psi_k(r)e^{-i\omega_k^Rt }
e^{-\gamma\omega_k^I t}
\end{equation}
 Projecting on to the states $|e\rangle$ and $|g\rangle$
 and integrating over spatial coordinates in each case we arrive at the
equations governing the time development of the amplitudes $(C_g$,
and $C_e)$;
\begin{eqnarray}
i\dot{C_g}=C_g(t)(\omega_g^R-i\gamma\omega_g^I)+C_e(t)
\Omega_R\overrightarrow{V}(t)e^{-\gamma\omega_a^It}e^{-i\omega_a^Rt}\\
i\dot{C_e}=C_e(t)(\omega_e^R-i\gamma\omega_e^I) +
C_g(t)\Omega_R\overrightarrow{V}(t)e^{\gamma\omega_a^It}e^{-i\omega_a^Rt}
\end{eqnarray}
 where the Rabi
frequency is defined as $\Omega_R = \frac{eE_0}{\hbar}\langle
e|r|g\rangle$ and the dipole approximation has been used.

Introducing the transformations $\tilde{C_g}=[C_g
e^{(i\omega_g^R+\gamma\omega_g^I)t}]$  and $\tilde{C_e}=[C_e
e^{(i\omega_e^R+\gamma\omega_e^I)t}]$
\begin{eqnarray}
i\tilde{C_g}=\tilde{C_e}\Omega_R\overrightarrow{E}(t)e^{(-i\omega_a^R-\gamma\omega_a^I)t}\\
i\tilde{C_e}=\tilde{C_g}\Omega_R\overrightarrow{E}(t)e^{(i\omega_a^R+\gamma\omega_a^I)t}
\end{eqnarray}
$\overrightarrow{E}(t)= \frac {e^{i\omega_L t}+e^{-i\omega_L
t}}{2}\;; \;
  \omega_L+\omega_a  =\omega_+ \; ;
\; \omega_L-\omega_a=\ \delta $ detuning frequency \\
Invoking the rotating wave approximation we get
\begin{eqnarray}
i\dot{\tilde{C_g}}=\frac{\Omega_R}{2}C_e
e^{-\gamma\omega_a^It}e^{i\delta
t}\\
i\dot{\tilde{C_e}}=\frac{\Omega_R}{2}C_g
e^{\gamma\omega_a^It}e^{-i\delta t}
\end{eqnarray}
Equation (26) and (27) can be uncoupled by the standard route,
leading to
\begin{eqnarray}
\ddot{\tilde{C_g}}-(\gamma\omega_a^I-i\delta )\dot{\tilde{C_g}}+\
\frac{\Omega_R^2}{4}\tilde{C_g}=0 \\
\ddot{\tilde{C_e}}+(\gamma\omega_a^I-i\delta)\dot{\tilde{C_e}}+\
\frac{\Omega_R^2}{4}\tilde{C_e}=0
\end{eqnarray}
Taking the initial condition that $C_g(0)=1 \; ;C_e(0)=0$ we get the
solutions
\begin{eqnarray}
\tilde{C_g}&=& e^{-\frac{\Delta}{2}t}\left[\cos\frac{\Omega_{Rd}
t}{2}+
i \frac{\Delta}{\Omega_{Rd}} \sin \frac{\Omega_{Rd} t}{2}\right]\\
\tilde{C_e}&=& e^{\frac{\Delta}{2}t}\;\left[
i\frac{\Omega_R}{\Omega_{Rd}} \sin\frac{\Omega_{Rd}t}{2}\right]
\end{eqnarray}
where $\Delta=\gamma\omega_a^I-i\delta$     $\;$ and $\;$
$\Omega_{Rd}=\sqrt{\Omega_R^2-{(\gamma\omega_a^I-i\delta)}^2}$.\\
Hence, \begin{eqnarray}
C_e=e^{-\gamma(\frac{\omega_g^I+\omega_e^I}{2})t}e^{-i\frac{\delta}{2}t}e^{-i\omega_e^Rt}\left[
i\frac{\Omega_R}{\Omega_{Rd}} \sin\frac{\Omega_{Rd}t}{2}\right]
\end{eqnarray}
Hence, the excited state population $P_e=|C_e|^2$ is given by
\begin{eqnarray}
|C_e|^2=e^{-\gamma(\omega_g^I+\omega_e^I)t}\left[
\frac{{\Omega_R}^2}{{\Omega_{Rd}}^2}
\sin^2\frac{\Omega_{Rd}t}{2}\right]
\end{eqnarray}
  The result shows that the contribution of a given state to the
evolving wave function $(\Psi)$ of the system at a particular time
is given in terms of the damping coefficient and the sum of the
energies of the two levels coupled by laser light. The coherent
temporal oscillations of the population in the excited state
obtained above matches with the experimental observations made by Yu
et al \cite{Yu02} in Jopsepshon phase qubit. The observed
oscillatory behaviour of the decaying amplitude reported by them is
successfully explained by our model based on the description of a
damped quantum dot by a non-hermitian Hamiltonian in real
Coordinates. The probability of being in the state 'k' (e or g) at
any given time is therefore given by $P_k(t)=|C_k(t)|^2$. For the
excited state 'e' Figure 2 shows the nature of the time dependence
of $P_e$. As expected it is coherently oscillatory and exponentially
damped.We note that the equation 33 was earlier developed by Yu et
al \cite{Yu02} as the asymtotic limit of solution of the appropriate
Lioville equation for the density operator under the rotating wave
approximation, and used to interpret their experimental observation.
We have arrived at the same results based on the non-hermitian
Hamiltonian.

\section{Conclusion}
In summary the proposed model describes correctly the effects of
inherent damping in  a quantum dot. The amount of dissipating energy
in a particular state in a quantum dot is naturally related to the
damping coefficient. The decoherence of Rabi oscillations shows that
the rate of decoherence is exponentially related not only to the
damping coefficient but also to the energy separation between the
two levels. Again one interesting point is that the inherent life
times of all the different states is predictable assuming that the
life time of any one particular state are known from experiment. We
also note that the temporal coherent oscillation of population in
Josephson junction is correctly explained by the present model.
\section{Appendix}
In atomic units the Hamiltonian of equation (14) reads
\begin{eqnarray}
H=-\frac{1}{2m_e}\left(\frac {\partial^2}{\partial
r^2}+\frac{1}{r}\frac{\partial}{\partial
r}+\frac{1}{r^2}\frac{\partial^2}{\partial\phi^2}\right)-
\frac{i\gamma}{m_e}\left(1+r\frac{\partial}{\partial r}\right)-\frac
{i\omega_c}{2}\left(\frac{\partial}{\partial\phi}\right)+\Omega_d^2r^2
\end{eqnarray}
where $\Omega_d^2= \frac{1}{2}m_e\left[\frac
{\omega_c^2}{4}+\frac{\gamma^2}{m_e^2}+\omega_0^2\right]$.\\
Substituting, $f(r,\phi)=\frac{e^{(il\phi)} \Psi(r)}{\sqrt{2\pi}}$
and multiplying both sides by
$\sqrt{2\pi}e^{-il\phi}r^{\frac{1}{2}}$ leads to radial Schrodinger
equation,
\begin{eqnarray}
-\frac{1}{2m_e}\left[\frac{
d^2}{dr^2}+\frac{1}{4r^2}-\frac{l^2}{r^2} + \Omega_d^2r^2\right]
f(r)+\frac{lw_{c}}{2}-\frac{i\gamma}{m_e}\left\{ f(r)+ r
f'(r)-\frac{1}{2}f(r)\right\} = E f(r)
\end{eqnarray}
or
\begin{eqnarray}
\left[\frac{ d^2}{dr^2}+(\frac{1}{4}-l^2)
\frac{1}{r^2}-\Omega_d'^2r^2-m_e\omega_cl +
2i\gamma(\frac{1}{2}+r\frac{\partial}{\partial r}) + 2Em_e
\right]f(r) = 0
\end{eqnarray}
Where
$\Omega_d'^2=m_e^2(\frac{\omega_c^2}{4}+\frac{\gamma^2}{m_e^2}+\omega_0^2)$\\
Substituting $r=\frac{x}{\sqrt{\Omega_d'}}$ the radial function
$f(r)$ changes to $g(x)$.
\begin{eqnarray}
\left[\frac{ d^2}{dx^2} + \left(\frac{1}{4}-l^2\right) \frac{1}{x^2}
- m_e\frac{(\omega_cl-2E)}{\Omega_d'} + 2i\frac{\gamma}
{\Omega_d'}\left(\frac{1}{2}+x\frac{d}{d x}\right)+2Em_e\right]g(x)
=0
\end{eqnarray}
Asymptotic analysis leads to \\
$g(x)=g_0(x)V(x)g_\infty(x)$ ;
$g_0(x)=e^{-x^2/2}$,  $g_\infty(x)=x^{\frac{1}{2}+|l|}$\\
$g(x)=e^{-x^2/2} V(x) x^{\frac{1}{2}+|l|}$.\\
Where $V(x)= \sum_{j}b_{j}x^{j}$ satisfies laguure series.
 Again taking $z=x^2$ the function $V(x)$ changes to function q(z), such
 that $q(z) = \sum_{k}a_{k}z^{k} $ satisfy the Laguure series.
\begin{eqnarray}
\nonumber [z\frac{d^2}{dz^2}+(l+1-z)\frac{d}{dz}-\{\frac{l+1}{2}+
m_e\frac{(\omega_cl-2E)}{4\Omega_d'}\}\\+2i\frac{\gamma}{4}(1++2z\frac{d}{d
z}+l-z)]q(z) =0
\end{eqnarray}\\
where E is complex $(=E_R-i\Gamma)$. The Lageurre series satisfies
equation (36). Accordingly the total wave function reads
\begin{eqnarray} \psi_{n,l}(r, \phi)=\frac
{C}{2\sqrt{\pi}}e^{i\phi}e^{-\frac {\Omega_d^2 r^2}{2}}r^l
\sum_{0}^{n} b^n r^{2n}
\end{eqnarray}
\section{Acknowledgement}
M. Mukhopadhyay  would like to thank the CSIR of Goverment of India,
 New Delhi for the award senior research fellowship. S.P.
 Bhattacharyya thanks the DST for a generous research grant.

\newpage
\begin{figure}
\begin{center}
\includegraphics[width=8cm]{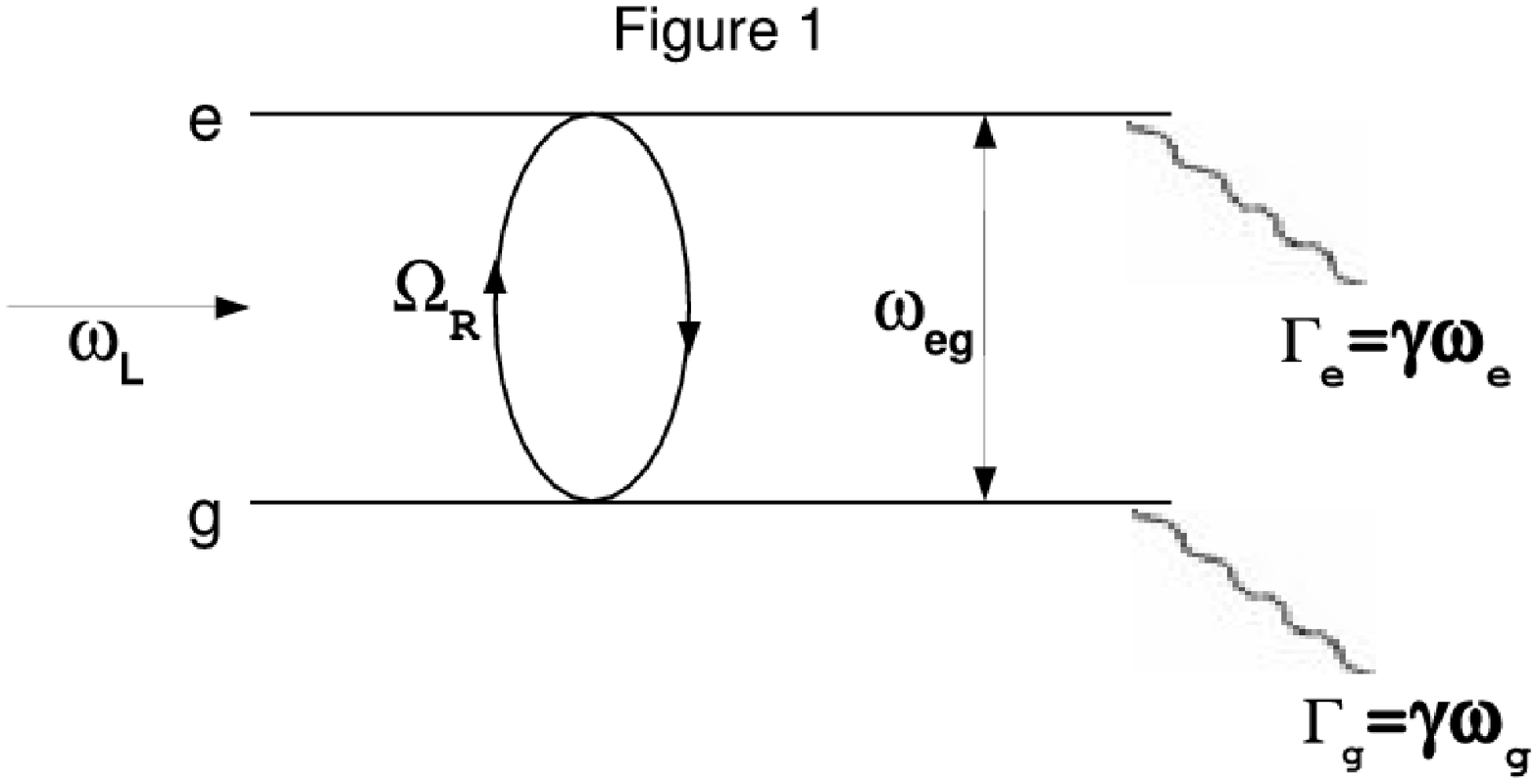}
\caption {Energy level diagram for two-level system showing decay rates for ground and excited
states $\Gamma_g$, $\Gamma_e$ respectively. The excited and ground state frequency difference is denoted by
$\omega_{eg}$. The Rabi oscillation $\Omega_R$ is introduced by an
external laser frequency $\omega_L$.}
\end{center}
\end{figure}
\begin{figure}
\centerline{\psfig{file=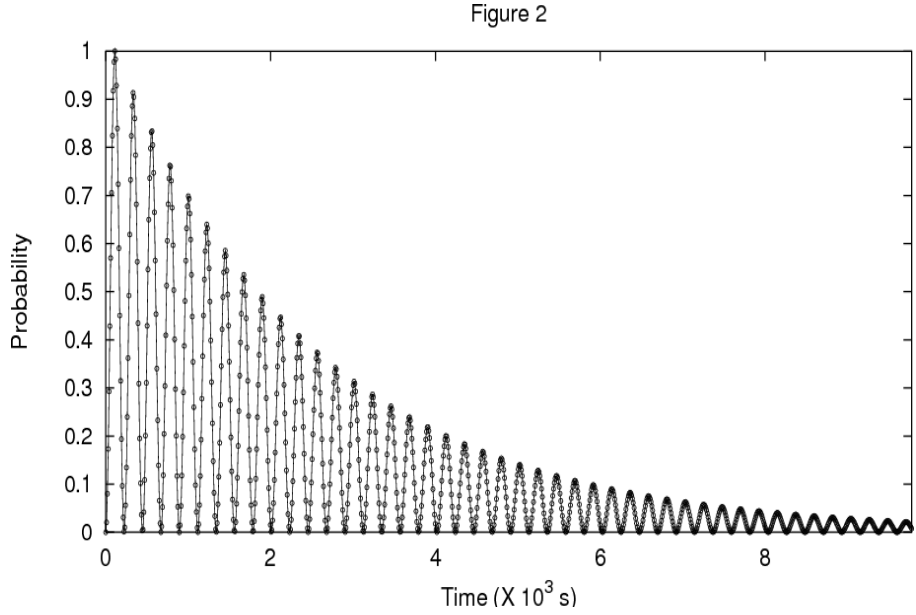,width=12cm,height=8cm,angle=270}}
\caption{A plot of total population of the excited states versus
real time (in a.u) with $m_e=$1 a.u., $\omega_c=10^{-3}$ a.u., $\omega_0=$0.0141
a.u., $\Omega_R=0.01$ a.u., $(\omega_g+\omega_e)=0.04$ a.u. and $\gamma=0.0001$ a.u.}
\end{figure}

\end{document}